\documentclass[twocolumn,aps,prl,superscriptaddress]{revtex4-2}   
\usepackage[utf8]{inputenc}
\usepackage{color}
\usepackage{colortbl}
\usepackage{bm}
\usepackage{amsmath}
\usepackage{amssymb}
\usepackage{graphicx}
\usepackage[pdfusetitle, bookmarks=true,bookmarksnumbered=true,bookmarksopen=true,bookmarksopenlevel=1, breaklinks=true,pdfborder={0 0 1},backref=false,colorlinks=true] {hyperref}
\hypersetup{ pdfborderstyle=}

\makeatletter
\newcommand{\lyxmathsym}[1]{\ifmmode\begingroup\def\b@ld{bold}
  \text{\ifx\math@version\b@ld\bfseries\fi#1}\endgroup\else#1\fi}

\usepackage[caption=false]{subfig}
\makeatother

\begin{document}
\title{Macroscopic Coherent Axion Production by Reverse Parametric Fluorescence}

\author{Zhan Bai}
\affiliation{State Key Laboratory of Ultraintense Laser Science and Technology, 
Shanghai Institute of Optics and Fine Mechanics, 
Chinese Academy of Sciences, China}

\author{Baifei Shen}
\affiliation{ShanghaiTech University, China}
\affiliation{Shanghai Normal University, China}

\author{Liangliang Ji}
\email{jill@siom.ac.cn}
\affiliation{State Key Laboratory of Ultraintense Laser Science and Technology,
Shanghai Institute of Optics and Fine Mechanics, 
Chinese Academy of Sciences, China}

\affiliation{ShanghaiTech University, China}
\author{Ruxin Li}
\affiliation{State Key Laboratory of Ultraintense Laser Science and Technology,
Shanghai Institute of Optics and Fine Mechanics, 
Chinese Academy of Sciences, China}
\affiliation{ShanghaiTech University, China}

\date{\today}

\begin{abstract}
We propose a macroscopically coherent laboratory source of axion-like particles (ALPs) 
through the axion--electron coupling \(g_{ae}\). 
Two counterpropagating optical modes drive reverse parametric fluorescence in a crystal, 
where two pump photons are converted into a relativistic ALP through virtual ionic transitions, while the medium returns to its initial state. 
Phase matching enables emission amplitudes from many ions to add coherently without preparing material coherence. 
The pump frequencies determine the ALP energy, making the source continuously tunable. 
The production rate scales with the product of the two pump powers and the square of the source length.
Resonant absorption followed by fluorescence completes the detection scheme. 
For benchmark crystal and laser parameters, a one-year operation gives a reach of \(g_{ae}\simeq2.8\times10^{-11}\),
substantially improving the sensitivity of purely laboratory-based searches for low-mass ALPs.
\end{abstract}
\maketitle

\begin{figure*}[t]
\centering 
\subfloat[\label{fig:energy-levels}]{\includegraphics[width=0.205\textwidth]{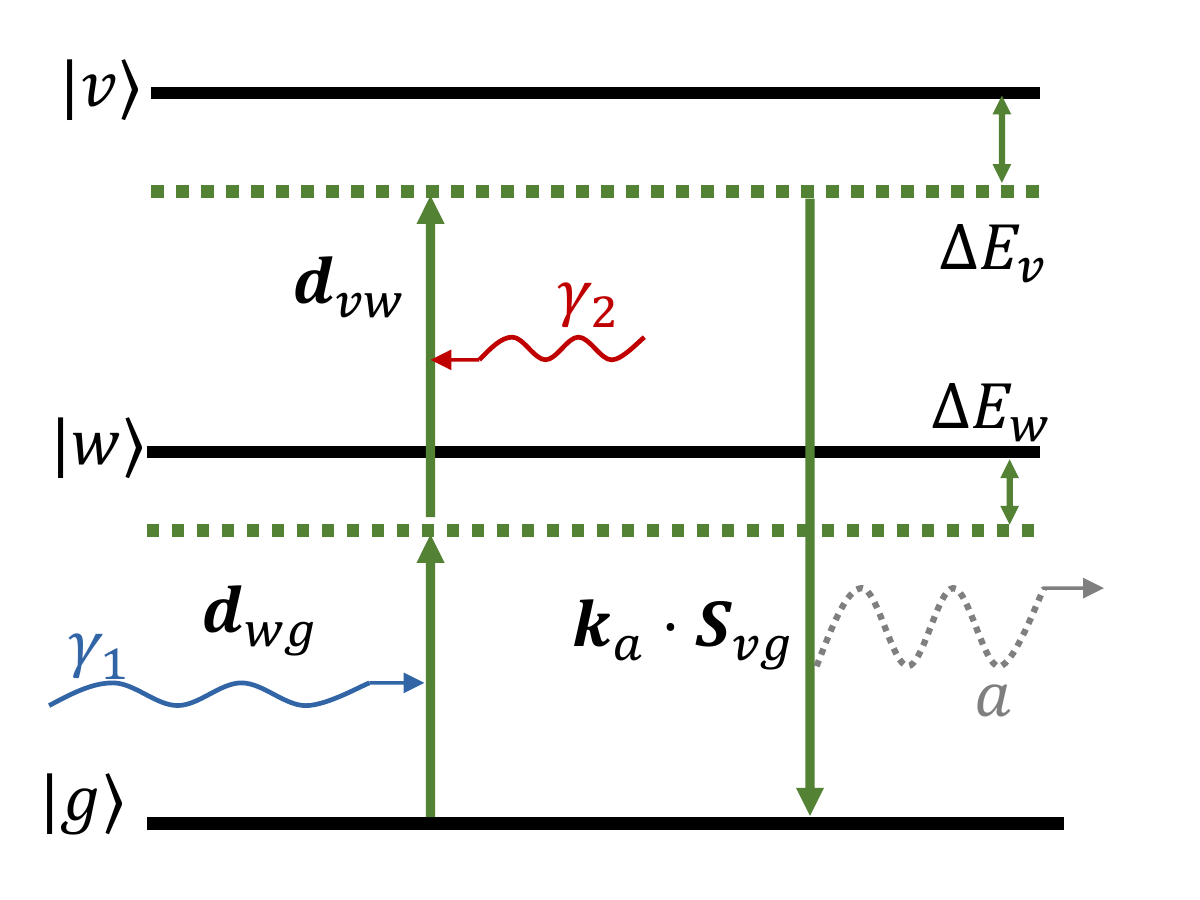} }
\subfloat[\label{fig:scheme}]{\includegraphics[width=0.57\textwidth]{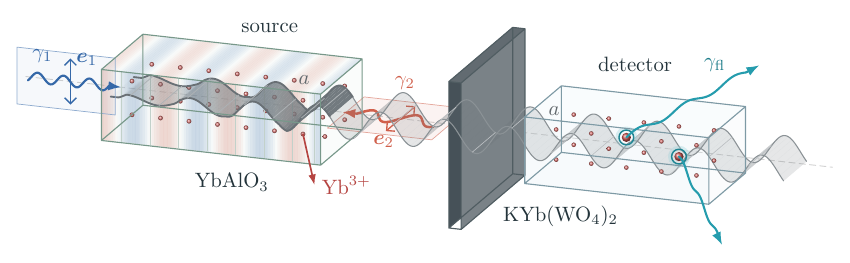} }
\subfloat[\label{fig:det-levels}]{\includegraphics[width=0.205\textwidth]{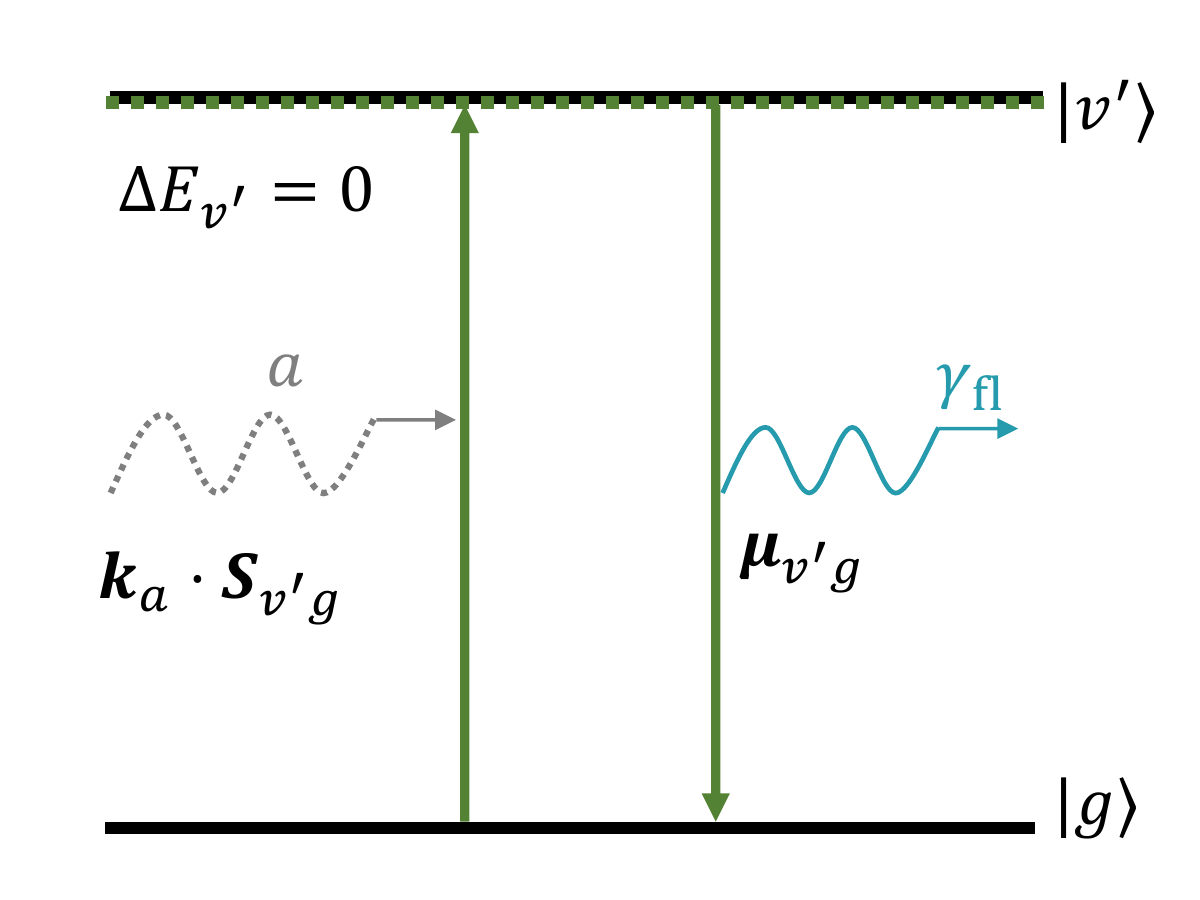} }
\caption{Source--detector mechanism. (a) Two optical vertices and one
ALP spin vertex form a virtual single-ion cycle. (b) The phase-matched
choice uses nondegenerate counterpropagating pumps to produce a forward
ALP beam. (c) The ALP resonantly excites a real detector transition
followed by fluorescence. Phase matching is required only in the source.}
\end{figure*}

\textit{Introduction}---Laboratory production of axions and axion-like particles (ALPs) through the axion--electron coupling \(g_{ae}\)
has been pursued through electron bremsstrahlung and secondary electromagnetic showers in fixed-target and beam-dump experiments
~\citep{Bechis:1979kp,Riordan:1987aw,NA64:2021aiq,NA64:2021xzo,CCM:2021jmk,Capozzi:2023ffu,Eberhart:2025lyu,LDMX:2026mrx},
Compton-like conversion \(\gamma+e^-\to a+e^-\) in reactor cores
~\citep{Dent:2019ueq,AristizabalSierra:2020rom,NEON:2024kwv,Mirzakhani:2025bqz},
and the recently proposed ALP radiation from electrons in intense laser fields
~\citep{Vacalis:2025uzk,He:2025jyo,Chen:2025gar}.
Because different electrons have no common ALP-emission phase, 
their contributions add incoherently, 
so the total rate grows only linearly with the number of electrons \(N\).

Coherence enables a qualitatively distinct production mechanism. 
It has already been proposed that forward scattering from atoms
provides coherent photon-to-ALP conversion through \(g_{ae}\) 
~\citep{Flambaum:2018wbu}.
For a single-photon channel, however, macroscopic buildup requires both
\(\omega_{\gamma}=\omega_a\) and \(\boldsymbol{k}_{\gamma}=\boldsymbol{k}_a\).
In a dielectric, \(|\boldsymbol{k}_{\gamma}|=n_r\omega_{\gamma}\) with
\(n_r>1\), whereas a relativistic on-shell ALP has \(k_a\simeq\omega_a\);
the resulting momentum mismatch prevents coherence over a macroscopic source length. 

Two-photon kinematics can remove this obstruction, 
because the two photon frequencies and wave vectors can be tuned independently. 
This principle underlies paired superradiance (PSR), 
a macrocoherent two-photon process proposed for cosmic axion detection~\citep{Huang:2019rmc,Sasao:2017zdn}. 
However, it requires a prepared material coherence. 
Its finite lifetime limits the available interaction time, and the high-gain regime may be sensitive to vacuum-seeded startup~\citep{Bai:2026uyd}.

The key observation of this work is that parametric fluorescence provides a different form of macroscopic phase coherence, 
one mediated entirely by virtual material transitions. 
In ordinary parametric fluorescence, 
a pump photon is converted into a photon pair through virtual transitions in a medium, 
which returns to its initial state after each event~\citep{Grynberg:2010QO}.
This mechanism has recently been used for probing the cosmic neutrino background~\citep{Huang:2025yqu}.
As no real material coherence needs to be prepared or sustained, 
the process is not bounded by the lifetime of an excited collective state.

In this work, we propose to use the reverse process, 
in which two highly occupied optical modes combine in medium 
and emit an ALP through the axion--electron coupling \(g_{ae}\), 
while the medium returns to its initial state. 
Energy conservation fixes the ALP frequency, \(\omega_a=\omega_1+\omega_2\), 
while the vector phase-matching condition reads \(\boldsymbol{k}_1+\boldsymbol{k}_2=\boldsymbol{k}_a\). 
Under these conditions, the emission amplitudes from many ions add coherently, 
producing an \(N^{2}\) structure in the differential production rate. 

Microscopically, each ion undergoes the virtual cycle
\(|g\rangle\xrightarrow{\gamma_1}|w\rangle\xrightarrow{\gamma_2}|v\rangle\xrightarrow{a}|g\rangle\),
together with the exchange of \(\gamma_1\leftrightarrow\gamma_2\).
The one-photon detunings from \(|w\rangle\) are large. 
By contrast, the two-photon detuning from \(|v\rangle\) is chosen small enough to enhance the production amplitude, 
but still sufficiently large that \(|v\rangle\) remains off resonance. 
Thus both intermediate states are only virtually occupied.
We consider collinear counter-propagating modes so that \(|\boldsymbol{k}_{1}|-|\boldsymbol{k}_{2}|\approx \omega_a\) can match the momentum of a relativistic axion.
This construction requires the ion system to contain a state \(|v\rangle\) with the same parity as \(|g\rangle\) 
and an intermediate state \(|w\rangle\) with the opposite parity, 
allowing two electric-dipole optical couplings and an ALP-induced transition between \(|v\rangle\) and \(|g\rangle\) . 
As a concrete benchmark, we consider \(\mathrm{Yb}^{3+}\), 
whose level structure provides suitable candidate states.

For ALP detection, we employ a separate resonant-fluorescence mechanism.
Resonant absorption followed by optical readout has previously been proposed for cosmic axion detection
~\citep{Sikivie:2014lha,Braggio:2017oyt,Arvanitaki:2017nhi}.
Those schemes search for a narrow dark-matter line whose  frequency is fixed by the unknown particle mass, 
and therefore require tunable resonances or multiple transition systems to scan the relevant frequency range. 
In the present setup, by contrast, the ALP energy is fixed by the source itself and is continuously tunable through the two pump frequencies.
The detector can therefore be designed to operate on a chosen material resonance.

We implement this idea by exploiting host-dependent crystal-field shifts of the same active ion,
such as \(\mathrm{Yb}^{3+}\). 
The transition is kept slightly off resonance in the source crystal, 
ensuring that the production process proceeds through virtual intermediate states.
In a second crystal host,  
the corresponding transition \(|g\rangle\to|v'\rangle\) is shifted into resonance with the emitted ALP energy. 
An incident ALP then drives the real absorption process \(a+|g\rangle\to|v'\rangle\),
followed by radiative decay \(|v'\rangle\to|g\rangle+\gamma_{\mathrm{fl}}\).
The fluorescence photon provides a direct counting signature of ALP absorption.

We combine the phase-matched reverse-parametric-fluorescence source with this resonant-fluorescence detector. 
For fixed optical pump energies, the setup is sensitive to relativistic ALPs in the sub-meV mass range. 
With the benchmark crystal and laser parameters considered below, 
an estimated one-year run reaches a reference sensitivity of \(g_{ae}\simeq2.8\times10^{-11}\), 
extending the reach of purely laboratory-based searches for low-mass ALPs.

\textit{ALP Fluorescence Generation}---The level scheme for reverse parametric fluorescence into an ALP is shown in Fig.~\ref{fig:energy-levels}.
The relevant interactions are the electric-dipole coupling to the optical fields and the derivative ALP--electron spin coupling,
\[
\mathcal H_{\gamma}=-\bm d\cdot\bm E, \qquad \mathcal H_{ae}=\frac{g_{ae}}{m_e}\bm\nabla a\cdot\bm S_e.
\]

For an ion at \(\bm R_\ell\), 
third-order perturbation theory in the far-detuned regime gives the contribution 
mediated by an intermediate state \(|w\rangle\):
\begin{equation} \label{eq:main-counter-single-ion-amplitude}
\mathcal{M}_{\ell}(\bm{k}_a)=
\frac{g_{ae}}{m_{e}}\frac{\sqrt{\mathcal{W}_{1}\mathcal{W}_{2}}}{A n_r}
\frac{d^{2}_{\mathrm{eff}}S_{\mathrm{eff}}k_{a}}{E_{vg}-\omega_a}
\mathcal{T}_{12}\mathrm{e}^{\mathrm{i}\Delta\bm{k}\cdot\bm{R}_{\ell}},
\end{equation}
\[
\mathcal{T}_{12}= \frac{1}{2}\left[
\frac{(\hat{\bm{d}}_{vw}\cdot\hat{\bm{e}}_{2})(\hat{\bm{d}}_{wg}\cdot\hat{\bm{e}}_{1})}{E_{wg}-\omega_1}
+\frac{(\hat{\bm{d}}_{vw}\cdot\hat{\bm{e}}_{1})(\hat{\bm{d}}_{wg}\cdot\hat{\bm{e}}_{2})}{E_{wg}-\omega_2}
\right],
\]
where \(\mathcal W_{1,2}\) are the circulating powers of the two pump modes,
\(A\) is the transverse area and \(n_r\) is the refractive index. 
We also define
\(d_{\mathrm{eff}}^2=|\bm d_{vw}||\bm d_{wg}|\),
\(\hat{\bm d}_{ab}=\bm d_{ab}/|\bm d_{ab}|\),
\(S_{\mathrm{eff}}=|\hat{\bm k}_a\cdot\bm S_{gv}|\), 
and \(\Delta\bm{k} =\bm k_{\gamma 1}+\bm k_{\gamma 2}-\bm k_a\), 
with \(|\bm k_{\gamma 1,2}|=n_r\omega_{1,2}\). 
The vectors \(\hat{\bm e}_{1,2}\) are the polarization unit vectors of the two pump modes.
The two terms in \(\mathcal{T}_{12}\) correspond to the two photon-absorption time orderings, 
which generally have unequal one-photon detunings.

 \textit{Macroscopic coherence}---The total amplitude is the coherent sum over all ions, which for a uniform
ion density \(\rho\) becomes
\(\sum_\ell\mathcal M_\ell\to
\rho\int_V\mathrm{d}^3\bm R\,\mathcal M(\bm R)\).
Using the position-dependent phase
\(\exp(\mathrm{i}\Delta\bm k\cdot\bm R)\) in
Eq.~\eqref{eq:main-counter-single-ion-amplitude}, the spatial integral
factorizes as
\[
\mathcal I_\perp
=\frac{1}{A}\int_A\mathrm{d}^2\bm x_\perp\,
\mathrm{e}^{\mathrm{i}\Delta\bm k_\perp\cdot\bm x_\perp},
\qquad
\mathcal I_\parallel
=\int_0^{L_{\mathrm{src}}}\mathrm{d}z\,
\mathrm{e}^{\mathrm{i}\Delta k_z z}.
\]
The product \(|\mathcal I_\perp\mathcal I_\parallel|\) is maximal at
\(\Delta\bm k_\perp=0\) and \(\Delta k_z=0\), where all ions contribute
with the same phase. The temporal integral separately enforces energy
conservation.

We choose counterpropagating collinear pump modes, 
for which we have
\(\omega_a=\omega_1+\omega_2\) and 
\(k_a=k_{\gamma_1}-k_{\gamma_2}\)
For \(k_{\gamma_j}=n_r\omega_j\) and a relativistic ALP,
\(k_a\simeq\omega_a\), this gives
\begin{equation}
\omega_{1,2}
=\frac{1}{2}\left(\omega_a\pm\frac{\omega_a}{n_r}\right),
\label{eq:main-common-index-frequencies}
\end{equation}
The ALP energy can therefore be tuned by varying the pump frequencies while maintaining phase matching,
and the choice of \(\omega_a\) is determined by the detection material,
which will be discussed later.

The total production rate follows by integrating
\(|\mathcal M|^2\) over the ALP phase space, with a delta function
enforcing energy conservation:
\begin{equation}
\begin{aligned}
\Gamma_{\mathrm{src}}(m_a)
= \frac{g_{ae}^2\rho^2 k_a^3 d_{\mathrm{eff}}^4S_{\mathrm{eff}}^2}{8\pi^2m_e^2|\omega_a-E_{vg}|^2}
\frac{\mathcal W_1\mathcal W_2}{n_r^2}
|\mathcal{T}_{12}|^2
\mathcal F_{V},
\end{aligned}
\label{eq:main-source-rate}
\end{equation}
where \(\mathcal{F}_V\) is the finite-volume form factor defined by the remaining angular integral,
\begin{equation} \label{eq:main-counter-angular-integral}
\mathcal F_{V}(m_a)
=\int_{k_{az}>0}\mathrm{d}\Omega_a\,
|\mathcal I_\perp|^2|\mathcal I_\parallel|^2
\simeq \frac{4\pi^2 L_{\mathrm{src}}^2}{Ak_a^2},
\end{equation}
The last expression applies to a massless ALP at exact axial phase matching 
when the two transverse dimensions are comparable and sufficiently large.
The production rate then scales as
\(
\Gamma_{\mathrm{src}}
\propto
g_{ae}^2\mathcal W_1\mathcal W_2L_{\mathrm{src}}^2
\).
The detailed derivations are given in App.~\ref{app:axion-generation}.

\begin{figure}[!htpb]
\centering
\includegraphics[width=\columnwidth]{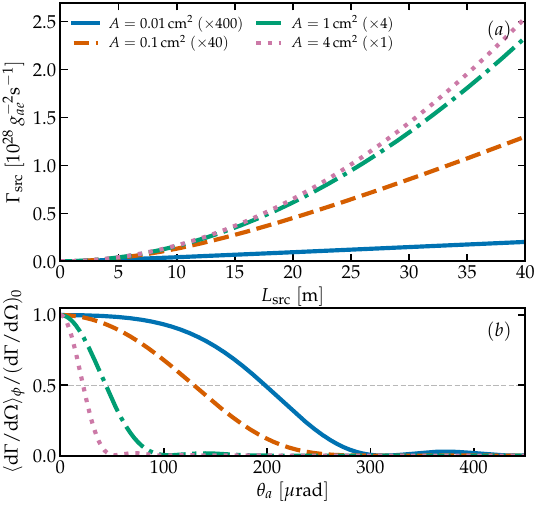}
\caption{ Signatures of finite-volume coherence in ALP production.
(a) The ALP production-rate coefficient as a function of source length;
the plotting multipliers are given in the legend.
(b) The corresponding azimuthally averaged differential rates,
normalized to their on-axis values, at \(L_{\mathrm{src}}=20\,\mathrm{m}\).
Both panels use \(m_a=0\), exact axial phase matching, and a fixed circulating
intensity of \(160\,\mathrm{kW\,cm^{-2}}\) in each beam.
The increasing angular collimation with \(A\) accompanies the extension of the
quadratic length-scaling regime in panel (a), displaying the same transverse
coherent site sum.
}
\label{fig:source-area-length-scaling}
\end{figure}

Although each ion emits an outgoing spherical ALP wave, 
phase matching makes the ions across the source cross section act collectively as a phased emitter. 
Their interference confines the emission to an angle \(\theta_a\sim(k_a\sqrt A)^{-1}\), 
so appreciable transverse spreading occurs only beyond \(L_{\mathrm{crit}}\sim k_aA\), 
of the order of a Rayleigh length.
A transversely matched detector within this distance therefore intercepts most of the forward ALP flux. 
The complete integral retains both transverse diffraction and the mass-dependent longitudinal phase mismatch.
For \(L_{\mathrm{src}}\ll L_{\mathrm{crit}}\), 
the quadratic scaling is the direct signature that amplitudes from the full source length add coherently.
For \(L_{\mathrm{src}}\gg L_{\mathrm{crit}}\), 
angular integration suppresses the interference between longitudinal intervals separated by more than \(L_{\mathrm{crit}}\). 
These intervals then add effectively incoherently, 
so the total rate grows only linearly with \(L_{\mathrm{src}}\). 
Fig.~\ref{fig:source-area-length-scaling} displays the manifestations o finite-volume coherence.
Panel (a) shows the \(L_{\mathrm{src}}^2\) dependence at large \(A\).
At fixed \(L_{\mathrm{src}}=20\,\mathrm{m}\),
panel (b) shows the corresponding azimuthally averaged angular profiles,
normalized to their on-axis values.
Increasing \(A\) narrows the forward cone while extending the quadratic regime.

 \textit{Resonant fluorescence detection}---For detection we do not use the inverse channel \(a\to\gamma+\gamma\), 
which lacks the pump-power enhancement \(\propto\mathcal W_1\mathcal W_2\) of the driven source. 
Instead, the incident ALP is absorbed on a real ionic transition in a second material, 
and the excitation is read out through fluorescence (Fig.~\ref{fig:det-levels}):
\begin{equation}\label{eq:main-fluorescence-process}
a+|g\rangle\rightarrow|v'\rangle,\qquad|v'\rangle\rightarrow|g\rangle+\gamma_{\mathrm{fl}},
\end{equation}
where \(|v'\rangle\) is a real excited level at energy \(E_{v'g}\) above the ground state \(|g\rangle\).
The pump frequencies can be chosen via Eq.~\eqref{eq:main-common-index-frequencies} 
to bring \(\omega_a\) onto any selected detector line.

For the selected level \(|v'\rangle\), 
the weak-drive optical Bloch equations give the single-ion absorption cross section
\begin{equation}\label{eq:main-fluorescence-cross-section}
\sigma_{a}=\left(\frac{g_{ae}}{m_{e}}\right)^{2}\frac{k_{a}S^{2}_{v'g}\gamma_{2}}{(\omega_{a}-E_{v'g})^{2}+\gamma^{2}_{2}},
\end{equation}
where \(S^{2}_{v'g}\) is the projected spin factor of the transition 
and \(\gamma_{2}\) the coherence HWHM of the \(g\)--\(v'\) line.
The detailed derivation is given in App.~\ref{sec:resonant-fluorescence-detector}.

We adopt the independent-ion, weak-drive limit 
used in atomic-transition searches~\citep{Sikivie:2014lha,Braggio:2017oyt}. 
The fluorescence signal counts the population deposited in the selected level, 
so the single-ion absorption probabilities add along the ALP path.
The controlled ALP beam is taken to be much narrower than the material line,
\(\Delta\omega_a\ll\gamma_2\), 
and is tuned so that \(|\omega_a-E_{v'g}|\ll\gamma_2\). 
On resonance, the weak-absorption probability 
through an active-ion column of density \(\rho'\) and total length \(L_{\mathrm{det}}\) is
\begin{equation}
P_{\mathrm{det}}\simeq \rho'L_{\mathrm{det}}\sigma_{a}
=\rho'L_{\mathrm{det}}g^{2}_{ae}
\frac{k_{a}S^{2}_{v'g}}{m^{2}_{e}\gamma_{2}}.
\label{eq:main-fluorescence-probability}
\end{equation}
The detector response therefore depends only on the active-ion column density
\(\rho'L_{\mathrm{det}}\) and is unchanged if the detector is segmented at
fixed total column density.

Taking the produced ALPs to traverse the active detector column and using
the low-mass, exact-matching approximation in
Eq.~\eqref{eq:main-counter-angular-integral},
Eqs.~\eqref{eq:main-source-rate} and
\eqref{eq:main-fluorescence-probability} give
\[
N_{\mathrm{sig}}\simeq
\frac{g_{ae}^4}{m_e^4}\frac{T L_{\mathrm{src}}^{2}L_{\mathrm{det}} 
\mathcal W_{1}\mathcal W_{2}}
{2A n_{r}^{2}}\,
\frac{k_{a}^{2}d_{\mathrm{eff}}^{4}
\rho^{2}\rho'
S_{\mathrm{eff}}^{2}S_{v'g}^{2}}
{\gamma_{2}|E_{v'g}-E_{vg}|^{2}}\,
|\mathcal{T}_{12}|^{2}.
\]
The yield is quartic in \(g_{ae}\), quadratic in the coherent source
length, and linear in the detector length. This scaling directly reflects
the asymmetric source--detector architecture. 

\textit{Material Consideration}---A suitable ion must provide two same-parity levels separated by the target ALP energy, 
an opposite-parity state for the two virtual E1 vertices.
This property usually exists in rare-earth elements.
For a concrete estimate, we choose Yb\(^{3+}\) as a representative, 
whose \(4f^{13}\) configuration provides the same-parity ground \(^{2}F_{7/2}\) 
and excited \(^{2}F_{5/2}\) manifolds 
associated with \(|g\rangle\) and \(|v\rangle\), respectively. 
Their separation is approximately \(1.2\,\mathrm{eV}\), 
while charge-transfer (CT) excitations provide the opposite-parity virtual intermediate state \(|w\rangle\).

For the source, we consider \(\mathrm{YbAlO}_3\) (YbAP), 
with \(E_{vg}=1.26712\,\mathrm{eV}\) and \(E_{wg}=5.5\,\mathrm{eV}\)
~\citep{Boulon:2008JOSAB:25.884,vanPieterson:2000JLum:91.177}.
For the detector, we consider KYb(WO\(_4\))\(_2\) (KYbW), 
which has \(E_{v'g}=1.26315\,\mathrm{eV}\)~\citep{Pujol:2002PhRvB:65.165121}.
Fixing \(\omega_a=E_{v'g}\), the source detuning is
\(E_{vg}-\omega_a=3.97\,\mathrm{meV}\).
This is about five times the measured linewidth of the corresponding Yb:YAP transition~\citep{Song:2020OMExp:10.1522}, 
so we can treat
\(|v\rangle\) as off-resonant and virtual in the source.
The reported KYbW zero-line width gives
the resolution-limited coherence HWHM
\(\gamma_2=0.425\,\mathrm{meV}\)~\citep{Pujol:2002PhRvB:65.165121}.

The active-ion densities are
\(\rho=2.00\times10^{22}\,\mathrm{cm}^{-3}\) in YbAP and
\(\rho'=6.4\times10^{21}\,\mathrm{cm}^{-3}\) in KYbW
~\citep{Buryy:2010JPCM:22.055902,Pujol:2002PhRvB:65.165121}. 
For the source estimate, we approximate YbAP as an isotropic,
nondispersive medium with \(n_r=1.95\).
The corresponding pump frequencies are determined by Eq.~\eqref{eq:main-common-index-frequencies}.

The remaining microscopic matrix elements have not been measured for the
selected crystal-field branches and are estimated as detailed in
App.~\ref{app:material}. A free-ion angular reduction gives the reference
values \(S_{\mathrm{eff}}^{2}=S_{v'g}^{2}=1/7\). Using the available
charge-transfer lifetime and emission spectra, we estimate
\(d_{\mathrm{eff}}\simeq2.49\times10^{-5}\,\mathrm{eV}^{-1}\)
~\citep{Shim:2004RadM:38.493,Ricci:2011OptMa:33.1000}. The dipole directions and their
relative phase remain unknown.

For the reference estimate, we take \(x\parallel a\), \(y\parallel b\),
and \(\hat{\bm k}_a\parallel z\parallel c\). The four-site Pbnm sum
cancels for equal principal-axis polarizations and adds for crossed
polarizations; see App.~\ref{app:counter-site-selection}. We choose
\(\hat{\bm e}_1=\hat{\bm y}\) and
\(\hat{\bm e}_2=\hat{\bm x}\). Since
\(|E_{wg}-\omega_1|<|E_{wg}-\omega_2|\), the favorable orientation
\(\hat{\bm d}_{wg}\parallel\hat{\bm e}_1\) and
\(\hat{\bm d}_{vw}\parallel\hat{\bm e}_2\) gives, in the far-detuned
limit,
\begin{equation}
\mathcal T_{12}^{\max}
=\frac{1}{2|E_{wg}-\omega_1|},\qquad
|\mathcal T_{12}|
=|\eta|\mathcal T_{12}^{\max},
\label{eq:main-counter-T12-max}
\end{equation}
where \(|\eta|\leq1\).
Because the crossed channel is symmetry allowed and its normalized tensor contraction contains no parametrically small factor, 
its natural scale is \(|\eta|=O(1)\), 
while we take \(|\eta|=1\) for the optimistic benchmark. 
Its material-specific value requires polarization-resolved spectroscopy or first-principles calculations 
of the CT wave functions and transition tensors, 
and lies beyond the scope of this work.

\textit{Benchmark}---A schematic figure for the complete source--detector approach is shown in Fig.~\ref{fig:scheme}.
We place the source crystal in a resonant cavity 
and assume a common power buildup \(\mathcal B_{\mathrm{cav}}\) for the two pump modes. 
As a representative value, we take the ALPS II design buildup 
\(\mathcal B_{\mathrm{cav}}=1.6\times10^4\)~\citep{Ortiz:2020tgs}, 
with an external drive intensity \(I_1=I_2=10\,\mathrm{W\,cm^{-2}}\) 
and a transverse area \(A=4.0\,\mathrm{cm}^2\). 
We further take \(L_{\mathrm{src}}=L_{\mathrm{det}}=20\,\mathrm m\) and \(T=1\,\mathrm{yr}\). 
The circulating powers are \(\mathcal W_j=\mathcal B_{\mathrm{cav}}AI_j\), 
so the source combination \(\mathcal W_1\mathcal W_2/A\) in \(N_{\mathrm{sig}}\) 
becomes \(\mathcal B_{\mathrm{cav}}^{2}I_1I_2A\). 
Evaluating the exact finite-aperture calculation at \(m_a\to0\), 
the reference condition \(N_{\mathrm{sig}}=1\) gives
\begin{equation}
\begin{aligned}
g_{ae}^{\mathrm{lim}}\simeq&
2.8\times10^{-11}|\eta|^{-1/2}
\left(\frac{\mathcal B_{\mathrm{cav}}}{1.6\times10^4}\right)^{-1/2}\\
&\times
\left[\frac{I_1I_2}{(10\,\mathrm{W\,cm^{-2}})^2}\right]^{-1/4}
\left(\frac{A}{4.0\,\mathrm{cm}^2}\right)^{-1/4}\\
&\times\left(\frac{L_{\mathrm{src}}}{20\,\mathrm m}\right)^{-1/2}
\left(\frac{L_{\mathrm{det}}}{20\,\mathrm m}\right)^{-1/4}
\left(\frac{T}{1\,\mathrm{yr}}\right)^{-1/4}.
\end{aligned}
\label{eq:benchmark-reach-number}
\end{equation}
Eq.~\eqref{eq:benchmark-reach-number} separates the experimentally
adjustable scales from the material inputs fixed above. 

\begin{figure}[!t]
\centering
\includegraphics[width=0.94\columnwidth]{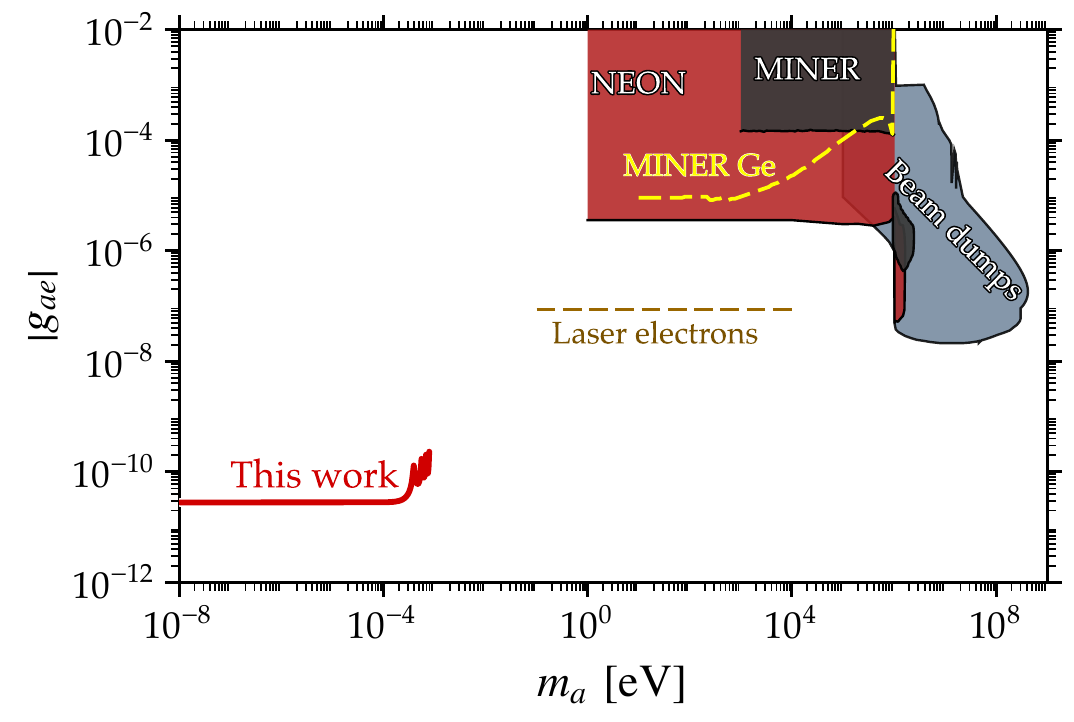}
\caption{Exclusion and projected \(g_{ae}\) reaches from this work and other pure laboratory-based approaches.
Filled regions show reactor exclusions from NEON~\citep{NEON:2024kwv}
and MINER~\citep{Mirzakhani:2025bqz}, together with the beam-dump envelope
formed from the historical beam-dump region used in the NEON comparison
and the modern pseudoscalar reanalysis of E137
~\citep{NEON:2024kwv,Eberhart:2025lyu}.
The MINER region contains its low-mass scattering branch and its separate MeV decay contour. 
Unfilled curves show the projected MINER Ge sensitivity ~\citep{Dent:2019ueq}, 
the laser-accelerated-electron sensitivity ~\citep{Vacalis:2025uzk}.
}
\label{fig:gae-reach} 
\end{figure}

Fig.~\ref{fig:gae-reach} shows the full mass dependence of \(g_{ae}^{\mathrm{lim}}\).
Exact phase matching gives a low-mass plateau near \(2.8\times10^{-11}\).
Above \(m_a\sim2\times10^{-4}\,\mathrm{eV}\),
longitudinal mismatch reduces the coherent source volume and the limit degrades.
We also present exclusions and projections from previous experiments and proposals.
Although the displayed portions of these projections lie in the eV--MeV range, 
the underlying reactor and laser schemes can be extended toward the \(m_a\to0\) limit
~\citep{Dent:2019ueq,NEON:2024kwv,Mirzakhani:2025bqz,Vacalis:2025uzk}.
The comparison highlights the sensitivity gained from macrocoherent production in the sub-meV regime.

\textit{Conclusion}---We have constructed a laboratory ALP source 
in which two highly occupied optical modes drive a virtual ionic cycle through \(g_{ae}\). 
Because every ion returns to the same state, 
phase matching allows their emission amplitudes to add across the source without preparing material coherence. 
In the collinear geometry considered here, two counterpropagating
pumps with unequal frequencies use their wave-vector difference to
compensate the dielectric momentum mismatch and generate a
forward-directed relativistic ALP beam.
At fixed transverse geometry, the source rate scales as
\(\Gamma_{\mathrm{src}}\propto g_{ae}^{2}\mathcal W_1\mathcal W_2L_{\mathrm{src}}^2\).
Resonant absorption followed by fluorescence closes the
source-and-counting experiment without an additional optical pump or
phase-matching requirement at the detector.

The principal significance is the coherent source itself: 
it turns the axion--electron interaction into a controllable laboratory production channel 
whose matched-mode yield receives a macroscopic coherent enhancement. 
For optimistic material parameters, 
the one-year benchmark reaches \(g_{ae}\simeq2.8\times10^{-11}\),
outperforming other laboratory-based projections shown. 
The precise numerical sensitivity depends on dedicated measurements of the relevant material properties.
The underlying macrocoherent enhancement, 
however, applies generally to any medium that supports the virtual ionic cycle 
and satisfies the phase-matching condition.
Our work establishes a novel route toward substantially larger laboratory ALP yields 
and source-based probes of the axion--electron coupling.

\textit{Acknowledgments}--This work is supported by the
National Natural Science Foundation of China (No. 12388102), the Strategic
Priority Research Program of the Chinese Academy of Sciences (No.
XDB0890000), the CAS Project for Young Scientists in Basic Research
(No. YSBR060). This work is also supported by the State Key Laboratory
of Dark Matter Physics.


%

\clearpage\onecolumngrid

\section*{Supplemental Materials}

\section{ALP generation from counterpropagating collinear beams} \label{app:axion-generation}

This appendix derives the source rate used in the main text for the counterpropagating collinear geometry. 
We first normalize the pump modes at fixed circulating power, 
and then derive the single-ion amplitude and its coherent sum over the finite crystal.

\subsection{Single-ion matrix element} \label{app:Single-Ion-Matrix-Element}

For each narrow-band traveling pump, the positive-frequency electric
field inside the crystal is
\begin{equation}
\hat{\boldsymbol{E}}^{(+)}_{j}=\mathrm{i}\sqrt{\frac{\omega_{j}v_{j}}{2n_rL}}\hat{\boldsymbol{e}}_{j}\frac{1}{\sqrt{A}}\hat{a}_{j}e^{-\mathrm{i}\omega_{j}t+\mathrm{i}k_{jz}z},\label{eq:app-counter-E-quantized}
\end{equation}
where \(k_{1z}=n_r\omega_1\), \(k_{2z}=-n_r\omega_2\),
\(v_j=(\mathrm{d}|k_{jz}|/\mathrm{d}\omega_j)^{-1}=n_r^{-1}\),
and \(A=L_xL_y\). A mode containing \(N_j\) photons carries circulating
power \(\mathcal W_j=v_jN_j\omega_j/L\), so that
\begin{equation}
\sqrt{N_j}\sqrt{\frac{\omega_jv_j}{2n_rLA}}
=\sqrt{\frac{\mathcal W_j}{2n_rA}}.
\label{eq:app-counter-fixed-power-normalization}
\end{equation}
The pumps occupy distinct frequency and polarization modes, so their
combined occupation factor is \(\sqrt{N_1N_2}\).

The microscopic source path contains two electric-dipole absorptions
and one derivative ALP vertex. Starting from 
\begin{equation}
\mathcal{L}_{ae}=\frac{g_{ae}}{2m_{e}}(\partial_{\mu}a)\bar{e}\gamma^{\mu}\gamma_{5}e,
\end{equation}
the leading nonrelativistic interaction Hamiltonians are 
\begin{equation}
\hat{H}_{ae}(t)=\frac{g_{ae}}{m_{e}}\boldsymbol{\nabla}\hat{a}(\boldsymbol{R}_{\ell},t)\cdot\hat{\boldsymbol{S}}_{\ell},\qquad\hat{H}_{\gamma}(t)=-\hat{\boldsymbol{d}}_{\ell}\cdot\hat{\boldsymbol{E}}(\boldsymbol{R}_{\ell},t).\label{eq:app-counter-interaction-Hamiltonians}
\end{equation}
Here \(\boldsymbol{R}_{\ell}\) is the ion position in the dipole approximation;
the velocity-suppressed time component of the axial current is omitted.

Let \(|g\rangle\) be the initial and final ionic state, \(|w\rangle\)
the retained opposite-parity CT level, and \(|v\rangle\) the
same-parity \(4f\) state connected to \(|g\rangle\) by the spin operator.
The initial and final states are 
\begin{equation}
|i\rangle=|g;N_{1},N_{2};0_{a}\rangle,\qquad|f\rangle=|g;N_{1}-1,N_{2}-1;1_{\boldsymbol{k}_{a}}\rangle.\label{eq:app-counter-initial-final-states}
\end{equation}
For the chronological sequence \(\gamma_{1}\to\gamma_{2}\to a\), the
three interaction-picture vertices are 
\begin{align}
\left\langle w;N_{1}-1,N_{2}\left|\hat{H}^{I}_{\gamma}(t_{1})\right|g;N_{1},N_{2}\right\rangle  & =-\mathrm{i}\sqrt{\frac{\mathcal{W}_{1}}{2n_rA}}(\boldsymbol{d}_{wg}\cdot\hat{\boldsymbol{e}}_{1})e^{\mathrm{i}(E_{wg}-\omega_{1})t_{1}+\mathrm{i}k_{1z}Z_{\ell}},\label{eq:app-counter-photon-vertex-1}\\
\left\langle v;N_{1}-1,N_{2}-1\left|\hat{H}^{I}_{\gamma}(t_{2})\right|w;N_{1}-1,N_{2}\right\rangle  & =-\mathrm{i}\sqrt{\frac{\mathcal{W}_{2}}{2n_rA}}(\boldsymbol{d}_{vw}\cdot\hat{\boldsymbol{e}}_{2})e^{\mathrm{i}(E_{v}-E_{w}-\omega_{2})t_{2}+\mathrm{i}k_{2z}Z_{\ell}},\label{eq:app-counter-photon-vertex-2}\\
\left\langle g;1_{\boldsymbol{k}_{a}}\left|\hat{H}^{I}_{ae}(t_{3})\right|v;0_{a}\right\rangle  & =-\mathrm{i}\frac{g_{ae}}{m_{e}}(\boldsymbol{k}_{a}\cdot\boldsymbol{S}_{gv})e^{\mathrm{i}(\omega_{a}-E_{vg})t_{3}-\mathrm{i}\boldsymbol{k}_{a}\cdot\boldsymbol{R}_{\ell}}.\label{eq:app-counter-axion-emission-vertex}
\end{align}

For the selected ordering, third-order perturbation theory gives 
\begin{equation}
S^{(3)}_{fi,12}=(-\mathrm{i})^{3}\int^{+\infty}_{-\infty}\mathrm{d}t_{3}\int^{t_{3}}_{-\infty}\mathrm{d}t_{2}\int^{t_{2}}_{-\infty}\mathrm{d}t_{1}\left\langle f\left|\hat{H}_{I}(t_{3})\hat{H}_{I}(t_{2})\hat{H}_{I}(t_{1})\right|i\right\rangle _{12}.\label{eq:app-counter-dyson-integral}
\end{equation}
Performing the ordered time integrals and replacing the infinitesimal
imaginary parts by the finite level widths gives
\(S^{(3)}_{fi,12}=2\pi\delta(\omega_a-\omega_1-\omega_2)
\mathcal M_{\ell,12}\), with
\begin{equation}
\mathcal{M}_{\ell,12}(\boldsymbol{k}_{a})=\frac{g_{ae}}{m_{e}}\frac{\sqrt{\mathcal{W}_{1}\mathcal{W}_{2}}}{2A n_r}
\frac{(\boldsymbol{k}_{a}\cdot\boldsymbol{S}_{gv})(\boldsymbol{d}_{vw}\cdot\hat{\boldsymbol{e}}_{2})(\boldsymbol{d}_{wg}\cdot\hat{\boldsymbol{e}}_{1})}
{\left(E_{vg}-\omega_{a}-\frac{\mathrm{i}}{2}\gamma_{v}\right)
\left(E_{wg}-\omega_{1}-\frac{\mathrm{i}}{2}\gamma_{w}\right)}
e^{\mathrm{i}[(k_{1z}+k_{2z})Z_{\ell}-\boldsymbol{k}_{a}\cdot\boldsymbol{R}_{\ell}]}.\label{eq:app-counter-one-ordering-M}
\end{equation}
The exchanged ordering follows from \(1\leftrightarrow2\). Define
\(\hat{\boldsymbol d}_{wg}=\boldsymbol d_{wg}/|\boldsymbol d_{wg}|\),
\(\hat{\boldsymbol d}_{vw}=\boldsymbol d_{vw}/|\boldsymbol d_{vw}|\), and
\(d_{\mathrm{eff}}^2=|\boldsymbol d_{wg}|\,|\boldsymbol d_{vw}|\).
The two orderings are then collected in
\begin{equation}
\mathcal{T}_{12}=\frac{1}{2}\left[
\frac{(\hat{\boldsymbol{d}}_{vw}\cdot\hat{\boldsymbol{e}}_{2})(\hat{\boldsymbol{d}}_{wg}\cdot\hat{\boldsymbol{e}}_{1})}
{E_{wg}-\omega_{1}-\frac{\mathrm{i}}{2}\gamma_{w}}
+\frac{(\hat{\boldsymbol{d}}_{vw}\cdot\hat{\boldsymbol{e}}_{1})(\hat{\boldsymbol{d}}_{wg}\cdot\hat{\boldsymbol{e}}_{2})}
{E_{wg}-\omega_{2}-\frac{\mathrm{i}}{2}\gamma_{w}}
\right].\label{eq:app-counter-single-pole-factor}
\end{equation}
Within the narrow forward lobe,
\(\boldsymbol k_a\cdot\boldsymbol S_{gv}\simeq k_aS_{\mathrm{eff}}\),
where \(S_{\mathrm{eff}}=|\hat{\boldsymbol c}\cdot\boldsymbol S_{gv}|\).
The spatial mismatch is
\begin{equation}
\boldsymbol{q}=(k_{1z}+k_{2z})\hat{\boldsymbol{z}}-\boldsymbol{k}_{a}
=-\boldsymbol{k}_{a\perp}+(k_{1z}+k_{2z}-k_{az})\hat{\boldsymbol{z}}.
\label{eq:app-counter-mismatch-vector}
\end{equation}
Adding the two time orderings gives the complete single-ion matrix
element
\begin{equation}
\mathcal{M}_{\ell}(\boldsymbol{k}_{a})=
\frac{g_{ae}}{m_{e}}\frac{\sqrt{\mathcal{W}_{1}\mathcal{W}_{2}}}{A n_r}
\frac{d^{2}_{\mathrm{eff}}\mathcal{T}_{12}}
{E_{vg}-\omega_{a}-\frac{\mathrm{i}}{2}\gamma_{v}}
S_{\mathrm{eff}}k_{a}e^{\mathrm{i}\boldsymbol{q}\cdot\boldsymbol{R}_{\ell}}.
\label{eq:app-counter-one-ion-amplitude}
\end{equation}
The Pbnm site sum and favorable estimate of \(\mathcal T_{12}\) are
given in App.~\ref{app:counter-site-selection}.

\subsection{Phase Matching and Production Rate} \label{app:counter-coherent-amplitude}

Every ion begins and ends in the same internal state, so no material
record identifies the emitting site. For a uniform active-ion density
\(\rho\), the phase factors in Eq.~\eqref{eq:app-counter-one-ion-amplitude}
therefore add at the amplitude level. Replacing the site sum by the
continuum integral gives
\begin{equation}
\sum_{\ell}\frac{1}{A}e^{\mathrm{i}\boldsymbol{q}\cdot\boldsymbol{R}_{\ell}}
\longrightarrow
\rho\mathcal{I}_{\perp}(\boldsymbol{k}_{a\perp})
\mathcal{I}_{\parallel}(k_{az}).
\label{eq:app-counter-coherent-sum}
\end{equation}
For a rectangular source, define
\(\Delta k_z=k_{1z}+k_{2z}-k_{az}\). The two volume factors are
\begin{align}
\mathcal{I}_{\perp}(\boldsymbol{k}_{a\perp})
&=\frac{1}{A}\int_A\mathrm{d}^2\boldsymbol{x}_{\perp}
e^{-\mathrm{i}\boldsymbol{k}_{a\perp}\cdot\boldsymbol{x}_{\perp}}
\nonumber\\
&=\text{sinc}\!\left(\frac{k_{ax}L_x}{2}\right)
\text{sinc}\!\left(\frac{k_{ay}L_y}{2}\right),
\label{eq:app-counter-Iperp}\\
\mathcal{I}_{\parallel}(k_{az}) & =\int^{L_{\mathrm{src}}}_{0}\mathrm{d}z\,
e^{\mathrm{i}\Delta k_{z}z}\nonumber \\
&=L_{\mathrm{src}}e^{\mathrm{i}\Delta k_zL_{\mathrm{src}}/2}
\text{sinc}\!\left(\frac{\Delta k_zL_{\mathrm{src}}}{2}\right).
\label{eq:app-counter-Iparallel}
\end{align}
Here \(\text{sinc}(x)=\sin x/x\). The temporal integral enforces energy
conservation, while the two spatial integrals are maximal at zero
transverse and longitudinal mismatch. The central coherent ray therefore
satisfies
\begin{equation}
\omega_{1}+\omega_{2}=\omega_{a},\qquad\boldsymbol{k}_{a\perp}=0,\qquad k_{1z}+k_{2z}=k_{az}.\label{eq:app-counter-phase-match}
\end{equation}
For the forward collinear solution adopted here, a degenerate
counterpropagating pair has nearly vanishing longitudinal momentum
difference and cannot match a relativistic ALP. The benchmark therefore
uses nondegenerate pumps.

The coherent matrix element of the finite source is therefore
\begin{equation} \label{eq:app-counter-coherent-amplitude}
  \mathcal{M}(\boldsymbol{k}_{a})=
\frac{g_{ae}\rho\sqrt{\mathcal{W}_{1}\mathcal{W}_{2}}}{m_{e}n_r}
\frac{d^{2}_{\mathrm{eff}}\mathcal{T}_{12}}
{E_{vg}-\omega_{a}-\frac{\mathrm{i}}{2}\gamma_{v}}
S_{\text{eff}}k_{a}\mathcal{I}_{\perp}\mathcal{I}_{\parallel}.
\end{equation}
At exact matching \(\mathcal{I}_{\parallel}=L_{\mathrm{src}}\). Squaring
Eq.~\eqref{eq:app-counter-coherent-amplitude} produces the macrocoherent
\(\rho^{2}L^{2}_{\mathrm{src}}\) factor. The transverse cross section
enters through \(\mathcal I_\perp\) and fixes the angular width of the
emitted ALP beam rather than supplying an independent coherent factor.

Using covariantly normalized one-ALP states, Fermi's golden rule gives
\begin{equation}
\Gamma_{\mathrm{src}}=\int\frac{\mathrm{d}^{3}\boldsymbol{k}_{a}}{(2\pi)^{3}2\omega_{a}}(2\pi)\delta(\omega_{a}-\omega_{1}-\omega_{2})|\mathcal{M}(\boldsymbol{k}_{a})|^{2}.\label{eq:app-counter-golden-rule}
\end{equation}
The energy delta function fixes
\(k_a=\sqrt{\omega_a^2-m_a^2}\), and the radial integral contributes
\(k_a/(8\pi^2)\). Together with the two powers of \(k_a\) from the
derivative ALP vertex, this gives the factor \(k_a^3\) below. Define the
finite-volume angular integral
\begin{equation}
\mathcal{F}_{V}(m_{a})=\int_{k_{az}>0}\mathrm{d}\Omega_{a}|\mathcal{I}_{\perp}|^{2}|\mathcal{I}_{\parallel}|^{2}.\label{eq:app-counter-FV}
\end{equation}
The total ALP production rate is
\begin{equation}\label{eq:app-counter-source-rate}
\Gamma_{\mathrm{src}}(m_{a})=
\frac{g_{ae}^2\rho^{2}\mathcal{W}_{1}\mathcal{W}_{2}
d^{4}_{\mathrm{eff}}S^{2}_{\text{eff}}k^{3}_{a}\mathcal{F}_{V}(m_{a})}
{8\pi^{2}m^{2}_{e}n_r^2
\left|E_{vg}-\omega_{a}-\frac{\mathrm{i}}{2}\gamma_{v}\right|^{2}}
|\mathcal{T}_{12}|^{2}.
\end{equation}
At exact matching and fixed transverse geometry,
\(\Gamma_{\mathrm{src}}\propto
g_{ae}^{2}\mathcal W_1\mathcal W_2L_{\mathrm{src}}^2\).

\subsection{Pbnm Polarization Selection and Optical Factor}
\label{app:counter-site-selection}

Each Pbnm translation cell of YbAlO$_3$ contains four
symmetry-related \(\mathrm{Yb}^{3+}\) ions. Their local transition-dipole
and spin matrix elements generally have different orientations, but
Pbnm symmetry relates them. We choose Cartesian axes parallel to the
crystal principal axes,
\begin{equation}
\hat{\boldsymbol{x}}\parallel\hat{\boldsymbol{a}},\qquad
\hat{\boldsymbol{y}}\parallel\hat{\boldsymbol{b}},\qquad
\hat{\boldsymbol{z}}\parallel\hat{\boldsymbol{c}},
\label{eq:app-counter-reference-site-axes}
\end{equation}
and take \(\hat{\boldsymbol{k}}_a\parallel\hat{\boldsymbol{z}}\), so
the ALP vertex selects \(S_z\).

Choose one of the four ions as the reference site and regard its
\(S_z\) and the \(x\) and \(y\) components of
\(\boldsymbol d_{wg}\) and \(\boldsymbol d_{vw}\) as given. For pump
1 polarized along \(r\) and pump 2 along \(s\), with
\(r,s\in\{x,y\}\), denote its single-ion matrix element by
\(\mathcal M_0^{(rs)}\). The part that determines its relative sign
under the site operations is
\begin{equation}
\mathcal M^{(rs)}_0\propto
S_z\left[
\frac{(\boldsymbol d_{vw})_s(\boldsymbol d_{wg})_r}
{E_{wg}-\omega_1-\frac{\mathrm{i}}{2}\gamma_w}
+
\frac{(\boldsymbol d_{vw})_r(\boldsymbol d_{wg})_s}
{E_{wg}-\omega_2-\frac{\mathrm{i}}{2}\gamma_w}
\right].
\label{eq:app-counter-reference-site-amplitude}
\end{equation}
The rotational parts of the Pbnm operations that generate the other
three sites are \(C_{2x}\), \(C_{2y}\), and \(C_{2z}\). For either
transition dipole, their action on the transverse components and
\(S_z\) is
\begin{align}
C_{2x}:&(d_x,d_y,S_z)\longrightarrow(+d_x,-d_y,-S_z),\nonumber\\
C_{2y}:&(d_x,d_y,S_z)\longrightarrow(-d_x,+d_y,-S_z),\nonumber\\
C_{2z}:&(d_x,d_y,S_z)\longrightarrow(-d_x,-d_y,+S_z).
\label{eq:app-counter-reference-site-rotations}
\end{align}
These are proper rotations, so spin and dipole components transform
with the same Cartesian signs. Equation~\eqref{eq:app-counter-reference-site-rotations}
therefore fixes the relative amplitude of every site once the
reference-site amplitude is specified.

For two pumps polarized along the same axis, each term in
Eq.~\eqref{eq:app-counter-reference-site-amplitude} contains two
dipole components with the same rotational sign. Their product is
unchanged, leaving the \(S_z\) signs \((+,-,-,+)\) for
\(\mathbb I,C_{2x},C_{2y},C_{2z}\), respectively. Hence
\begin{equation}
\mathcal M_{\mathrm{cell}}^{(rr)}
=\mathcal M_0^{(rr)}(1-1-1+1)=0,
\qquad r=x,y.
\label{eq:app-counter-reference-site-parallel-sum}
\end{equation}
For orthogonal polarizations, both time orderings contain one \(x\)
and one \(y\) dipole component. Under each rotation, the sign of their
product compensates any sign change of \(S_z\), and all four sites
therefore contribute in phase:
\begin{equation}
\mathcal M_{\mathrm{cell}}^{(rs)}
=\mathcal M_0^{(rs)}(1+1+1+1)
=4\mathcal M_0^{(rs)},
\qquad \{r,s\}=\{x,y\}.
\label{eq:app-counter-reference-site-orthogonal-sum}
\end{equation}
Thus the translation-cell sum eliminates the \(xx\) and \(yy\)
channels and permits the \(xy\) and \(yx\) channels. Since
\(x\parallel a\) and \(y\parallel b\), these are respectively the
forbidden \(aa\) and \(bb\) channels and the allowed \(ab\) and \(ba\)
channels. The fractional translations in the full Pbnm operations
contribute a phase \(e^{\mathrm{i}\boldsymbol q\cdot\boldsymbol t}\).
It equals unity on the phase-matched central ray and differs from
unity only by \(O(|\boldsymbol q|a_{\mathrm{cell}})\ll1\) across the
narrow coherent lobe, so it does not alter this selection rule.

The cross-polarized channel is realized with separate propagation
eigenmodes: pump 1 is polarized along \(y\parallel b\), and pump 2
along \(x\parallel a\). For the benchmark frequencies, the
\(\omega_1\) denominator in
Eq.~\eqref{eq:app-counter-single-pole-factor} has the smaller
magnitude. The favorable single-pole orientation is therefore
\begin{equation}
\hat{\boldsymbol d}_{wg}\parallel\hat{\boldsymbol e}_1,\qquad
\hat{\boldsymbol d}_{vw}\parallel\hat{\boldsymbol e}_2,
\label{eq:app-counter-reference-site-favorable-orientation}
\end{equation}
for which
\begin{equation}
\mathcal T_{12}^{\max}
=\frac{1}
{2\left|E_{wg}-\omega_1-\frac{\mathrm{i}}{2}\gamma_w\right|},
\qquad
|\mathcal T_{12}|=|\eta|\mathcal T_{12}^{\max}.
\label{eq:app-counter-reference-site-optimistic-factor}
\end{equation}
Because the crossed channel is symmetry allowed and its normalized tensor contraction is not parametrically suppressed, its natural scale is \(|\eta|=O(1)\). 
We therefore take \(|\eta|=1\) in the optimistic benchmark. A material-specific value would require polarization-resolved spectroscopy or first-principles calculations of the CT wave functions and transition tensors, which are beyond the scope of this work.

\section{Resonant ALP Absorption and Fluorescence Readout} \label{sec:resonant-fluorescence-detector}

At the detector, 
an incident ALP resonantly excites a selected real ionic transition. 
The excited ion subsequently decays, 
emitting a fluorescence photon that provides the observable signal. 
In this section, we derive this excitation probability.

We treat the incident ALP as classical wave
\begin{equation}
a(\boldsymbol{x},t)
=a_{0}\cos(\omega_{a}t-\boldsymbol{k}_{a}\cdot\boldsymbol{x}),
\qquad \omega_{a}^{2}=k_{a}^{2}+m_{a}^{2},
\label{eq:res-fl-classical-field}
\end{equation}

Consider the selected transition
\(\lvert g\rangle\to\lvert v'\rangle\), with
\(E_{v'g}=E_{v'}-E_{g}\) and
\(\boldsymbol{S}_{v'g}=\langle v'|\hat{\boldsymbol{S}}|g\rangle\).
The positive-frequency component of the ALP field is
\(a^{(+)}=(a_{0}/2)e^{-\mathrm{i}\omega_{a}t
+\mathrm{i}\boldsymbol{k}_{a}\cdot\boldsymbol{x}}\).
In the rotating-wave approximation, the single-ion Hamiltonian is
\[
\hat{H}_{0} =E_{v'g}|v'\rangle\langle v'|,\qquad
\hat{H}_{I} =V_{v'g}e^{-\mathrm{i}\omega_{a}t} |v'\rangle\langle g|+\text{h.c.},
\]
where the ALP-induced coupling amplitude is
\begin{align} \label{eq:res-fl-semiclassical-drive}
V_{v'g} =\mathrm{i}\frac{g_{ae}a_{0}}{2m_{e}} \boldsymbol{k}_{a}\cdot\boldsymbol{S}_{v'g} e^{\mathrm{i}\boldsymbol{k}_{a}\cdot\boldsymbol{R}}.
\end{align}
The ion position \(\boldsymbol{R}\) enters only through the phase of
\(V_{v'g}\), which cancels from the single-ion rate.

The quantum state of a single ion is described by the density operator
\[
\rho=\sum_{i,j\in\{g,v'\}}\rho_{ij}|i\rangle\langle j|,
\qquad
\rho_{ij}\equiv\langle i|\rho|j\rangle.
\]
The diagonal elements \(\rho_{gg}\) and \(\rho_{v'v'}\) give the populations of the ground and excited states, respectively, 
whereas the off-diagonal element \(\rho_{v'g}\) describes their quantum coherence. 
In the absence of relaxation, the density operator evolves according to the von Neumann equation,
\begin{equation}
\dot{\rho}=-\mathrm{i}
\left[\hat{H}_{0}+\hat{H}_{I}(t),\rho\right].
\label{eq:res-fl-von-neumann-equation}
\end{equation}

We work in a frame rotating at \(\omega_a\), 
reusing the same symbol for the transformed coherence,
\(e^{\mathrm{i}\omega_a t}\rho_{v'g}\longrightarrow\rho_{v'g}\).
Including population decay at rate \(\gamma_1\) and coherence decay at rate \(\gamma_2\), 
the Bloch equations are
\begin{align}
\dot{\rho}_{v'v'}
&=-\gamma_{1}\rho_{v'v'}
+\mathrm{i}\left(V^{*}_{v'g}\rho_{v'g}
-V_{v'g}\rho^{*}_{v'g}\right),
\label{eq:res-fl-bloch-population}\\
\dot{\rho}_{v'g}
&=-\left[\gamma_{2}+\mathrm{i}(E_{v'g}-\omega_{a})\right]
\rho_{v'g}
-\mathrm{i}V_{v'g}\left(\rho_{gg}-\rho_{v'v'}\right).
\label{eq:res-fl-bloch-coherence}
\end{align}

In the weak-drive limit, depletion and saturation are negligible, so
Eq.~\eqref{eq:res-fl-bloch-coherence} may be evaluated with
\(\rho_{gg}\simeq1\) and \(\rho_{v'v'}\simeq0\). Setting
\(\dot{\rho}_{v'g}=0\) in the steady state then gives
\begin{equation}
\rho_{v'g}
=-\frac{\mathrm{i}V_{v'g}}
{\gamma_{2}+\mathrm{i}(E_{v'g}-\omega_{a})}.
\label{eq:res-fl-steady-coherence}
\end{equation}
Setting \(\dot{\rho}_{v'v'}=0\) in
Eq.~\eqref{eq:res-fl-bloch-population} gives the population balance
\begin{equation}
\gamma_{1}\rho_{v'v'}
=\mathrm{i}\left(
V_{v'g}^{*}\rho_{v'g}
-V_{v'g}\rho_{v'g}^{*}\right)
=\Gamma_{a,g\to v'}.
\label{eq:res-fl-population-balance}
\end{equation}
Under the ideal-readout assumption, each radiative decay produces one
counted fluorescence photon, so
\(\Gamma_{\mathrm{fl}}=\gamma_{1}\rho_{v'v'}\). Combining
Eqs.~\eqref{eq:res-fl-steady-coherence} and
\eqref{eq:res-fl-population-balance} gives
\begin{align}
\Gamma_{\mathrm{fl}}(\omega_{a})
=\Gamma_{a,g\to v'}(\omega_{a})
&=\frac{2|V_{v'g}|^{2}\gamma_{2}}
{(E_{v'g}-\omega_{a})^{2}+\gamma_{2}^{2}}\nonumber\\
&=\frac{g_{ae}^{2}a_{0}^{2}k_{a}^{2}}{2m_{e}^{2}}
S_{v'g}^{2}
\frac{\gamma_{2}}
{(\omega_{a}-E_{v'g})^{2}+\gamma_{2}^{2}},
\label{eq:res-fl-single-ion-rate}
\end{align}
where
\(S_{v'g}^{2}=|\hat{\boldsymbol{k}}_{a}\cdot
\boldsymbol{S}_{v'g}|^{2}\).
This result agrees with the two-level absorption rate in Ref.~\citep{Arvanitaki:2017nhi}.

The single-ion absorption cross section is the excitation rate per unit
incident ALP number flux,
\(\sigma_a(\omega_a)\equiv\Gamma_{a,g\to v'}(\omega_a)/\Phi_a\).
For the classical wave in Eq.~\eqref{eq:res-fl-classical-field}, the
cycle-averaged energy density and corresponding number flux are
\begin{equation}
\langle\rho_{a}\rangle=\frac{1}{2}\omega_{a}^{2}a_{0}^{2},
\qquad
\Phi_{a}=\frac{\langle\rho_{a}\rangle}{\omega_{a}}
\frac{k_{a}}{\omega_{a}}
=\frac{1}{2}k_{a}a_{0}^{2}.
\label{eq:res-fl-classical-flux}
\end{equation}
Here \(\langle\rho_a\rangle/\omega_a\) is the ALP number density and
\(k_a/\omega_a\) is its group velocity. Combining
Eqs.~\eqref{eq:res-fl-single-ion-rate} and
\eqref{eq:res-fl-classical-flux} gives
\begin{align} \label{eq:res-fl-cross-section-general}
\sigma_{a}(\omega_{a}) =\frac{g_{ae}^{2}}{m_{e}^{2}}k_{a}S_{v'g}^{2}
\frac{\gamma_{2}}
{(\omega_{a}-E_{v'g})^{2}+\gamma_{2}^{2}},
\end{align}
On resonance, \(\omega_a=E_{v'g}\), the cross section reduces to
\begin{equation}
\sigma_{a}^{\mathrm{res}}
=\frac{g_{ae}^{2}k_{a}S_{v'g}^{2}}
{m_{e}^{2}\gamma_{2}}.
\label{eq:res-fl-cross-section-resonant}
\end{equation}
For independent active ions of number density \(\rho'\), the
weak-absorption probability is
\begin{equation}
P_{\mathrm{abs}}
\simeq \rho'L_{\mathrm{det}}\sigma_{a}^{\mathrm{res}},
\qquad
\rho'L_{\mathrm{det}}\sigma_{a}^{\mathrm{res}}\ll1.
\label{eq:res-fl-count-probability}
\end{equation}

\section{Material and atomic-level reduction} \label{app:material}

The source and detector both use Yb\(^{3+}\), but they use different
parts of the same atomic structure. This section first builds the
free-ion \(4f^{13}\) basis, then specifies the crystal-field-split
levels of YbAlO\(_{3}\)
and lowest effective CT pole used at the source, and finally gives
the real KYbW detector line. The spin factor is analytic, but the
two CT dipole directions and their relative phase are not presently
measured; they therefore remain in the explicit orientation factor
\(\eta\).

\subsection{Spin Reduction and \texorpdfstring{YbAlO\(_3\)}{YbAlO3} Source Levels} \label{app:spin}

According to  Ref.~\citep{Cowan:1981AtomSpectra}, 
the ground state \(|g\rangle\) of \(\textrm{Yb}^{3+}\) belongs to the \(4f^{13}\,{}^{2}F_{7/2}\) ground multiplet,
and \(|v\rangle\) belongs to the \(4f^{13}\,{}^{2}F_{5/2}\) excited multiplet, 
with free-ion quantum numbers \((L,S,J)=(3,1/2,7/2)\) and \((3,1/2,5/2)\), respectively.
The ALP interaction couples these states through the physical electron-spin operator \(\bm{S}\). 
The Wigner--Eckart theorem and the standard recoupling formula involving a Wigner \(6j\) symbol then give
\begin{equation}
\left\langle (LS)J'\Vert\bm{S}\Vert(LS)J\right\rangle =(-1)^{L+S+J'+1}\sqrt{(2J'+1)(2J+1)}\begin{Bmatrix}S & J' & L\\
J & S & 1
\end{Bmatrix}\sqrt{S(S+1)(2S+1)}.\label{eq:app-spin-reduced}
\end{equation}
For \(L=3\), \(S=1/2\), \(J=7/2\), and \(J'=5/2\), the corresponding
Wigner \(6j\) symbol is 
\begin{equation}
\left|\begin{Bmatrix}\frac{1}{2} & \frac{5}{2} & 3\\
\frac{7}{2} & \frac{1}{2} & 1
\end{Bmatrix}\right|=\frac{1}{\sqrt{21}},
\end{equation}
which gives 
\begin{equation}
\left|\left\langle ^{2}F_{5/2}\Vert\bm{S}\Vert{}^{2}F_{7/2}\right\rangle \right|^{2}=\frac{24}{7}.\label{eq:app-spin-number}
\end{equation}
For an unpolarized initial \(J=7/2\) manifold and an isotropic average
over the three components of the vector operator, the mean squared
projection is 
\begin{equation}
S^{2}_{\text{eff}}\equiv\frac{1}{3(2J+1)}\left|\left\langle ^{2}F_{5/2}\Vert\bm{S}\Vert{}^{2}F_{7/2}\right\rangle \right|^{2}=\frac{1}{7},\qquad S_{\text{eff}}=\frac{1}{\sqrt{7}}.\label{eq:app-Seff}
\end{equation}
This is a free-ion angular average, rather than a measured matrix element
for a specific crystal-field transition.

For the source benchmark, we next specify the YbAlO\(_3\) ion density
and the crystal-field-split levels associated with \(|g\rangle\) and \(|v\rangle\).
The reported lattice parameters of YbAlO\(_{3}\) are~\citep{Buryy:2010JPCM:22.055902}
\begin{equation}
a=5.1261\,\text{\AA},\qquad b=5.3314\,\text{\AA},\qquad c=7.3132\,\text{\AA}.
\end{equation}
Each Pbnm translation cell contains four Yb\(^{3+}\) ions, giving the
source-ion number density
\begin{equation}
\rho=\frac{4}{abc}=2.001\times10^{22}\,\mathrm{cm}^{-3}.\label{eq:app-ybap-density}
\end{equation}

Polarized spectroscopy of dilute Yb:YAP shows that the
crystal field splits the \(^{2}F_{7/2}\) and \(^{2}F_{5/2}\) manifolds
into four and three distinct energies, respectively~\citep{Boulon:2008JOSAB:25.884}:
\begin{equation}\label{eq:app-ybap-stark-data}
\begin{split}
^{2}F_{7/2}: & \quad0,\ 25.9,\ 42.3,\ 73.2\;\mathrm{meV}, \\
^{2}F_{5/2}: & \quad1.26712,\ 1.29068,\ 1.33035\;\mathrm{eV}.
\end{split}
\end{equation}

The detector benchmark fixes \(\omega_{a}=1.26315\,\mathrm{eV}\). We take
\(|g\rangle\) and \(|v\rangle\) to be the lowest levels of the
\(^{2}F_{7/2}\) and \(^{2}F_{5/2}\) manifolds, respectively, so that
\begin{equation}
\Delta E_{v}=E_{vg}-\omega_a=3.9675\,\mathrm{meV}.\label{eq:app-ybap-detuning}
\end{equation}

\subsection{Effective CT State and Electric-Dipole Input} \label{app:dipole}

In Yb-containing oxides, 
a charge-transfer excitation moves an electron from a neighboring O\(^{2-}\) ion to Yb\(^{3+}\), 
creating a high-energy state that can couple to the \(4f^{13}\) levels through electric-dipole transitions.
The CT absorption forms a broad band whose detailed level structure has not been resolved. 
For the benchmark, 
we represent its lowest contribution by a single effective state \(|w\rangle\) at \(E_{wg}=5.5\,\mathrm{eV}\), 
representative of Yb\(^{3+}\) CT absorption in oxide and aluminate hosts~\citep{vanPieterson:2000JLum:91.177}. 
The magnitudes of the two electric-dipole matrix elements are inferred below from the CT lifetime, 
while their directions and relative phase remain explicit in Eq.~\eqref{eq:app-counter-single-pole-factor}.

For a radiative branch \(w\to i\), with \(i=g,v\), where \(\hbar\omega_i\) and \(d_i\)
are the emitted photon energy and dipole magnitude, respectively, the
spontaneous-emission rate in the convention used here is~\citep{Loudon:2000QO}
\begin{equation}
\Gamma_{i}=\frac{n_{r}\omega^{3}_{i}|d_{i}|^{2}}{3\pi\epsilon_{0}\hbar c^{3}}.\label{eq:app-spontaneous}
\end{equation}

Pure YbAlO\(_{3}\) shows broad CT emission bands near \(3.65\) and \(2.48\,\mathrm{eV}\)~\citep{Ricci:2011OptMa:33.1000},
which we assign to decay from the same relaxed CT state into the \(^{2}F_{7/2}\) and \(^{2}F_{5/2}\) manifolds, respectively.
The \(5.5\,\mathrm{eV}\) pole energy is the vertical CT excitation entering
the virtual-state denominator, whereas the lower emission energies are
measured after lattice relaxation.
Their peak-energy separation, \(1.17\,\mathrm{eV}\), 
is close to \(E_{vg}=1.267\,\mathrm{eV}\); 
the \(0.10\,\mathrm{eV}\) difference reflects the use of broad-band maxima rather than resolved transition energies.
A low-temperature Yb:YAP sample series gives the CT lifetime
\(\tau_{\mathrm{CT}}=94.4\,\mathrm{ns}\)~\citep{Shim:2004RadM:38.493}.
For a branch fraction \(B_{i}\), \(\Gamma_{i}=B_{i}/\tau_{\mathrm{CT}}\), so 
\begin{equation}
|d_{i}|=\sqrt{\frac{3\pi\epsilon_{0}\hbar c^{3}B_{i}}{n_{r}\tau_{\mathrm{CT}}\omega^{3}_{i}}}.\label{eq:app-d-inversion}
\end{equation}

The branching fractions in YbAlO\(_3\) are unknown, so we take
\(B_g=B_v=1/2\) as an equal-branch reference.
Using \(n_{r}=1.95\) and converting the resulting dipole magnitudes to
natural units gives
\begin{equation}
\begin{aligned}
|\boldsymbol d_{wg}| & \simeq1.86\times10^{-5}\,\mathrm{eV}^{-1}, \qquad
|\boldsymbol d_{vw}| \simeq3.32\times10^{-5}\,\mathrm{eV}^{-1},\\
d_{\mathrm{eff}}&\equiv\sqrt{|\boldsymbol d_{wg}|\,|\boldsymbol d_{vw}|}
\simeq2.49\times10^{-5}\,\mathrm{eV}^{-1}.
\end{aligned}\label{eq:app-deff}
\end{equation}
More generally, 
\begin{equation}
d_{\mathrm{eff}}\simeq2.49\times10^{-5}\,\mathrm{eV}^{-1}\left(\frac{B_{g}B_{v}}{0.25}\right)^{1/4}\left(\frac{94.4\,\mathrm{ns}}{\tau_{\mathrm{CT}}}\right)^{1/2}.\label{eq:app-deff-scaling}
\end{equation}

\subsection{KYbW Detector Transition} \label{app:kybw-levels}

The detector uses one real transition within the same Yb\(^{3+}\) \(4f^{13}\)
configuration. In stoichiometric KYb(WO\(_{4}\))\(_{2}\), low-temperature
spectroscopy identifies a transition between the lowest levels of the
\(^{2}F_{7/2}\) and \(^{2}F_{5/2}\) manifolds at
\(1.26315\,\mathrm{eV}\)~\citep{Pujol:2002PhRvB:65.165121}.
At \(6\,\mathrm{K}\), the population is concentrated in the lowest ground
level, so the selected line fixes
\begin{equation}
\omega_{a}=E_{v'}-E_{g}=1.26315\,\mathrm{eV}.\label{eq:app-kybw-zero-line-energy}
\end{equation}

The selected line has a resolution-limited FWHM of \(0.85\,\mathrm{meV}\)
at \(1.26315\,\mathrm{eV}\), measured with an instrumental resolution of
\(0.64\,\mathrm{meV}\)~\citep{Pujol:2002PhRvB:65.165121}.
We use half of this full width, \(\gamma_{2}=0.425\,\mathrm{meV}\),
as the coherence HWHM, assuming the controlled ALP line is narrower.
For the detector spin matrix element, we use the free-ion estimate
\(S^{2}_{v'g}=1/7\) from Eq.~\eqref{eq:app-Seff}.

\end{document}